# Measuring Traffic

Peter J. Bickel, Chao Chen, Jaimyoung Kwon, John Rice, Erik van Zwet and Pravin Varaiya

*Abstract.* A traffic performance measurement system, PeMS, currently functions as a statewide repository for traffic data gathered by thousands of automatic sensors. It has integrated data collection, processing and communications infrastructure with data storage and analytical tools. In this paper, we discuss statistical issues that have emerged as we attempt to process a data stream of 2 GB per day of wildly varying quality. In particular, we focus on detecting sensor malfunction, imputation of missing or bad data, estimation of velocity and forecasting of travel times on freeway networks.

## 1. INTRODUCTION

As vehicular traffic congestion has increased, especially in urban areas, so have efforts at data collection, analysis and modeling. This paper discusses the statistical aspects of a particular effort, the Freeway Performance Measurement System (PeMS). We begin this introduction with some general discussion of data collection and traffic modeling and then describe PeMS.

### 1.1 Data Collection and Traffic Modeling

Traffic data are collected by three types of sensors. The first type is a point sensor, which provides estimates of flow or volume, occupancy and speed at a particular location on the freeway, averaged over 30 seconds. Ninety percent of point sensors are inductive loops buried in the pavement; the others are overhead video cameras or side-fired radar detectors. Point sensors provide continuous measurement. The large amount of data they provide can be used for statistical analysis.

The second type of sensors are implemented by floating cars that record GPS or tachometer readings from which one can construct the vehicle trajectory. Floating cars are expensive since they require drivers. Departments of Transportation (DoTs) typically deploy floating cars once or twice a year on stretches of freeway that are congested to determine travel time and the extent of the freeway that is congested. The data are insufficient for reliable estimates of travel time variability.

The third type of sensor can be used in areas in which vehicles are equipped with RFID tags. These tags are used for electronic toll collection (ETC). In the San Francisco Bay Area, for example, ETC tags are used for bridge toll collection. ETC readers are deployed at several locations, in addition to the bridge toll booths. These readers collect the tag ID and add a time stamp. By matching these at two consecutive reader locations, one gets the vehicle's travel time between the two locations. (One may view these data as samples of floating car trajectories.) The www.511.org site displays travel times estimated using these data. Of course, this type of sensor can only be deployed in a few locations. Moreover, the penetration of ETC tags in the whole vehicle population, and hence the data they provide, varies by time of day and day of week.

*Peter J. Bickel is is Professor, Department of Statistics, University of California, Berkeley, Berkeley, California 94720, USA e-mail: bickel@stat.berkeley.edu. Chao Chen is a graduate student, TFS Capital, 121 N. Walnut Street Ste 320, West Chester, Pennsylvania 19380, USA e-mail: chao@tfscapital.com. Jaimyoung Kwon is Assistant Professor, Department of Statistics, California State University, East Bay, Hayward, California 94542, USA e-mail: jaimyoung.kwon@csueastbay.edu. John Rice is Professor, Department of Statistics, University of California, Berkeley, Berkeley, California 94720, USA e-mail: rice@stat.berkeley.edu. Erik van Zwet is with the Mathematical Institute, University of Leiden, 2300 RA Leiden, The Netherlands e-mail: evanzwet@math.leidenuniv.nl. Pravin Varaiya is Nortel Networks Distinguished Professor, Department of Electrical Engineering and Computer Science, University of California, Berkeley, Berkeley, California 94720, USA e-mail: varaiya@eecs.berkeley.edu.*

In addition, there are special data sets obtained from surveys.

Point sensors implemented by inductive loops provide 95% of the data used by DoTs and traffic analysts worldwide. These data are used for two purposes: real-time traffic control and building traffic flow models for planning.

The primary traffic control mechanism is ramp metering, which controls the volume of traffic that enters the freeway at an on-ramp. The rate of flow depends on the density of traffic on the freeway, estimated from real-time loop data. Measurement, modeling and control are discussed in Papageorgiou (1983) and Papageorgiou et al. (1990), for example.

Real-time and historical data are also used to estimate and predict travel times. Travel time predictions are posted on the web and on changeable message signs on the side of the freeway. Attempts to process these data to estimate the occurrence of an accident have been unsuccessful, because of high false alarm rates.

Simulation models are used by regional transportation planners to predict changes in the pattern of traffic through a freeway network as a result of projected increase in demand or the addition of a lane or extension of a highway. The models are more frequently used to predict the impact of proposed shopping or housing development, or, in an operational context, to compare different alternatives to relieve congestion at some location. Microscopic models, such as TSIS/CORSIM, TRANSIMS, VISSIM and Paramics predict the movement of each individual vehicle. In macroscopic models, such as TRANSYT, SYNCHRO and DYNASMART, the unit of analysis is a platoon of vehicles or macroscopic variables such as flow, density and speed. URLs for these simulation models are given in the list of references. A fascinating overview and discussion of microscopic and macroscopic traffic models is provided by Helbing (2001).

Microscopic models are based on car-following and gap-acceptance models of driver behavior: how closely do drivers follow the car in front as a function of distance and relative speed; and how big a gap is needed before drivers change lanes. The parameters in these behavioral models are interpreted as indicators of driver aggressiveness and impatience. Microscopic models have scores of parameters, but they are calibrated using aggregate point detector data. As a result, most parameters are simply set to default values and no attempt is made to estimate them. Macroscopic models have fewer parameters, which can be estimated with point detector data. Typically, however, the estimates are based on least squares fit using a few days of data, with no attempt to calculate the reliability of the estimates. In order to predict network-wide traffic flows, the models need origin-destination flow data. These are converted into link-level flows assuming some kind of user equilibrium in which drivers take routes that have minimum travel times. Since these travel times depend on the link flows themselves, an iterative procedure is needed to calculate the assignment of origin-destination flows to link flows (Yu et al., 2004). Origin-destination flow data themselves are based on survey data or they are inferred from activity models that relate employment and household location data, obtained from the Census.

### 1.2 The Freeway Performance Measurement System

Over a number of years, the State of California has invested in developing Transportation Management Centers (TMCs) in urban areas to help manage traffic. The TMCs receive traffic measurements from the field, such as average speed and volume. These data, which are updated every 30 seconds, help the operations staff react to traffic conditions, to minimize congestion and to improve safety.

More recently, the California Department of Transportation (Caltrans) recognized that the data collected by the TMCs is valuable beyond real-time operations needs, and a concept of a central data repository and analysis system evolved. Such a system would provide the data to transportation stakeholders at all jurisdictional levels. It was decided to pursue this concept at a research level before investing significant resources. Thus, a collaboration between Caltrans and PATH (Partners for Advanced Transit and Highways) at the University of California at Berkeley was initiated to develop a performance measurement system or PeMS.

PeMS currently functions as a statewide repository for traffic data gathered by thousands of automatic sensors. It has integrated existing Caltrans data collection, processing and communications infrastructure with data storage and analytical tools. Through the Internet (http://pems.eecs.berkeley.edu), PeMS provides immediate access to the data to a wide variety of users. The system supports standard Internet browsers, such as Netscape or Explorer, so that users do not need any specialized software. In addition, PeMS provides simple plotting and analysis tools to facilitate standard engineering and planning tasks and help users interpret the data.

PeMS has many different users. Operational traffic engineers need the latest measurements to base their decisions on the current state of the freeway network. For example, traffic control equipment, such as ramp-metering and changeable message signs, must be optimally placed and evaluated. Caltrans managers want to quickly obtain a uniform and comprehensive assessment of the performance of their freeways. Planners look for long-term trends that may require their attention; for example, they try to determine whether congestion bottlenecks can be alleviated by improving operations or by minor capital improvements. They conduct freeway operational analyses, bottleneck identification, assessment of incidents and evaluation of advanced control strategies, such as on-ramp metering. Individual travelers and fleet operators want to know current shortest routes and travel time estimates. Researchers use the data to study traffic dynamics and to calibrate and validate simulation models. PeMS can serve to guide development and assess deployment of intelligent transportation systems (ITS).

PeMS has many different faces, but at some level it is just a simple balance sheet. A transportation system consumes public resources. In return, it produces transportation services that move people and goods. PeMS provides an automated system to account for these outputs and inputs through a collection of accounting formulas that aggregate received data into meaningful indicators. This produces a balance sheet for use in tracking performance over time and across agencies in a reasonably objective manner. Examples of "meaningful indicators" are:

- hourly, daily, weekly totals of VMT (vehicle-miles traveled), VHT (vehicle-hours traveled) and travel time for selected routes or freeway segments (links);
- means and variances of VMT, VHT and travel time.

These are simple measures of the volume, quality and reliability of the output of highway links. Publication each day of these numbers tells drivers and operators how well those links are functioning. Time series plots can be used to gauge monthly, weekly, daily and hourly trends.

Every 30 seconds, PeMS receives detector data over the Caltrans wide area network (WAN) to which all 12 districts are connected. Each individual Caltrans district is connected to PeMS through the WAN over a permanent ATM virtual circuit. A front end processor (FEP) at each district receives data from freeway loops every 30 seconds. The FEP formats these data and writes them into the TMC database, as well as into the PeMS database. PeMS maintains a separate instance of the database for each district. Although the table formats vary slightly across districts, they are stored in PeMS in a uniform way, so the same software works for all districts.

The PeMS computer at UC Berkeley is a four-processor SUN 450 workstation with 1 GB of RAM and 2 terabytes of disk. It uses a standard Oracle database for storage and retrieval. The maintenance and administration of the database is standard but highly specialized work, which includes disk management, crash recovery and table configuration. Also, many parameters must be tuned to optimize database performance. A part-time Oracle database administrator is necessary.

The PeMS database architecture is modular and open. A new district can be added online with six person-weeks of effort, with no disruption of the district's TMC. Data from new loops can be incorporated as they are deployed. New applications are added as need arises.

PeMS includes software serving three main functions: operating the database, processing and analyzing the data, and providing access to the data via the Internet. The processing of the data is done to ensure their reliability. It is a fact of life that the automatic detectors that generate most of our data are prone to malfunction. Detecting malfunction in an array of correlated sensors has been a statistical challenge. The related problem of imputation of bad or missing values is another major concern.

PeMS provides access to the database through the Internet. Using a standard browser such as Netscape or Internet Explorer, the user is able to query the database in a variety of ways. He or she can use built-in tools to plot the query results, or download the data for further study. Numerous tools for visualization are provided, allowing users to examine a variety of phenomena. Visualization tools include real-time maps showing levels of congestion, flow and speed profiles in space and in time, time series for individual detectors, plots displaying detector health, profiles of incidents in space and time, graphics to aid in the identification of bottlenecks, displays of delay as a function of space and time, and graphical summaries of vehicle miles traveled by freeway segment as a function of space and time.

In this paper we will describe how PeMS works. Our emphasis will be on the statistical issues that have emerged as we attempt to process a data stream of 2 GB per day of wildly varying quality. Real-time processing of the data is essential and while our methods cannot be optimal or "best" in any statistical sense, we aim for them to be as "good" as possible under the circumstances, and improvable over time.

The remainder of the paper is organized as follows. In Section 2 we describe the basic sensors upon which PeMS relies, loop detectors. In Section 3 we describe our approaches to detecting sensor malfunction and in Section 4 describe how we impute values that are missing or in error. Section 5 is devoted to a description of how we estimate velocity from the loop detectors, and Section 6 describes our method of predicting travel times for users. The reader will see that these efforts are very much a work in progress, with some aspects well developed and others under development.

## 2. LOOP DETECTORS

Caltrans TMCs currently operate many types of automatic sensors: microwave, infrared, closed circuit television and inductive loop. The most common type by far, however, is the inductive loop detector. Inductive loop detectors are wire loops embedded in each lane of the roadway at regular intervals on the network, generally every half-mile. They operate by detecting the change in inductance caused by the metal in vehicles that pass over them. A detector reports every 30 seconds the number of passing vehicles, and the percentage of time that it was covered by a vehicle. The number of vehicles is called *flow*, the percent coverage is called the *occupancy*. A roadside controller box operates a set of loop detectors and transmits the information to the local Caltrans TMC. This is done through a variety of media, from leased phone lines to Caltrans fiber optics. PeMS currently receives data from about 22,000 loop detectors in California.

A single inductance loop does not directly measure velocity. However, if the average length of the passing vehicles were known, velocity could be inferred from flow and occupancy. Estimation of velocity or, equivalently, average vehicle length has been an important part of our work, which is the subject of Section 5. At selected locations, two single-loop detectors are placed in close proximity to form a "double-loop" detector, which does provide direct measurement of velocity, from the time delay between upstream and downstream vehicle signatures. Most of the loop detectors in California are single-loop detectors while double-loop detectors are more widely used in Europe.

For a particular loop detector, the flow (volume) and occupancy at sampling time $t$ (corresponding to a given sampling rate) are defined as

$$(1) \quad q(t) = \frac{N(t)}{T}, \quad k(t) = \frac{\sum_{j \in J(t)} \tau_j}{T},$$

where $T$ is the duration of the sampling time interval, say 5 min, $N(t)$ is the number of cars detected during the sampling interval $t$, $\tau_j$ is the on-time of vehicle $j$, and $J(t)$ is the set of cars that are detected in time interval $t$. The traffic speed at time $t$ is defined as

$$v(t) = \frac{1}{N(t)} \sum_{j \in J(t)} v_j,$$

where $v_j$ is the velocity of vehicle $j$.

We will use $d, t, s, n$ to denote day, time of day, detector station and lane, letting them range over $1, \ldots, D$, $1, \ldots, T$, $1, \ldots, S$ and $1, \ldots, N$. By "station" we mean the collection of loop detectors in the various lanes at one location. Flow, occupancy, speed measured from station $s$, lane $l$ at time $t$ of day $d$ will be denoted as

$$q_{s,l}(d,t), k_{s,l}(d,t), v_{s,l}(d,t).$$

We will also index detectors by $i = 1, \ldots, I$ in some cases and use $t$ to denote sample times, so that notations like $q_i(t)$, $q_{s,l}(t)$, etc. will be seen as well.

Single-loop detectors are the most abundant source of traffic data in California, but loop data are often missing or invalid. Missing values occur when there is communication error or hardware breakdown. A loop detector can fail in various ways even when it reports values. Payne et al. (1976) identified various types of detector errors including stuck sensors, hanging on or hanging off, chattering, cross-talk, pulse breakup and intermittent malfunction. Even under normal conditions, the measurements from loop detectors are noisy; they can be confused by multi-axle trucks, for example.

Bad and missing samples present problems for any algorithm that uses the data for analysis, many of which require a complete grid of good data. Therefore, we need to detect when data are bad and discard them, and impute bad or missing samples in the data with "good" values, preferably in real time. The goal of detection and imputation is to produce a complete grid of clean data in real time.

## 3. DETECTING MALFUNCTION

Figure 1 illustrates detector failure. The figure shows scatter plots of occupancy readings in four lanes at a particular location. From these plots it can be inferred that loops in the first and second lanes suffer from transient malfunction.

The problem of detecting malfunctions can be viewed as a statistical testing problem, wherein the actual flow and occupancy are modeled as following a joint probability distribution over all loop detectors and times, and their measured values may be missing or produced in a malfunctioning state. Let $\Delta_i(t) = 0, 1, 2$ according as the state of detector $i$ at time $t$ is good, malfunctioning, or the data are missing. The problem of detecting malfunctioning is that of simultaneously testing $H : \Delta_i(t) = 0$ versus $K : \Delta_i(t) = 1$ or of estimating the posterior probabilities, $P(\Delta_i(t) = 1 | \text{data})$.

Since the model is too general and high dimensional for practical use, simplification is necessary. The most extreme and convenient simplification is to consider only the marginal distribution of individual (30-second) samples at an individual detector. In that case, the acceptance region and the rejection region partition the $(q, k)$ plane.

The early work in malfunction detection used heuristic delineations of this partition. Payne et al. (1976) presented several ways to detect various types of loop malfunctions from 20-second and 5-minute volume and occupancy measurements. These methods place thresholds on minimum and maximum flow, density and speed, and declare data to be invalid if they fail any of the tests. Along the same line, Jacobon, Nihan and Bender (1990) at the University of Washington defined an acceptable region in the $(q, k)$ plane, and declared samples to be good only if they fell inside. We will refer to this as the *Washington Algorithm.* This has an acceptance region of the form shown in Figure 2.

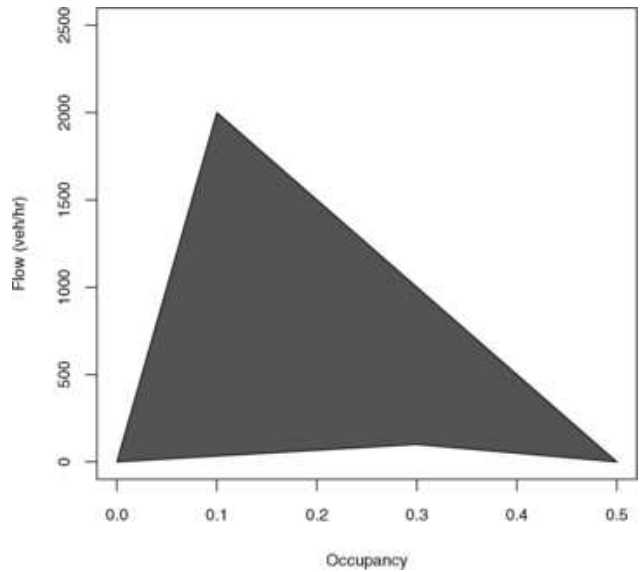

FIG. 2. Acceptance region of Washington algorithm.

PeMS currently uses a Daily Statistics Algorithm (DSA), proposed by Chen et al. (2003), which proceeds as follows. A detector is assumed to be either good or bad throughout the entire day. For day $d$, the following scores are calculated:

- $S_1(i, d) =$ number of samples that have occupancy $= 0$,
- $S_2(i, d) =$ number of samples that have occupancy $> 0$ and flow $= 0$,
- $S_3(i, d) =$ number of samples that have occupancy $> k^*$ $(= 0.35)$,
- $S_4(i, d) =$ entropy of occupancy samples $[-\sum_{x:p(x)>0} p(x) \log p(x)$ where $p(x)$ is the histogram of the occupancy]. If $k_i(d, t)$ is constant in $t$, for example, its entropy is zero.

Then the decision $\Delta_i = 1$ is made whenever $S_j > s_j^*$ for any $j = 1, \ldots, 4$. The values $s_j^*$ were chosen empirically. Since this algorithm does not run in real time, a detector is flagged as bad on the current day if it was bad on the previous day.

The idea behind this algorithm is that some loops seem to produce reasonable data all the time, while others produce suspect data all the time. Although it is very hard to tell if a single 30-second sample is good or bad unless it is truly abnormal, by looking at the time series of measurements for an entire day, one can usually easily distinguish bad behavior from good.

This procedure effectively corresponds to a model in which flow and occupancy measurement failures are independent and identically distributed across loops. The trajectory of detector $i$, $\{q_i(t); k_i(t); t = 1, \ldots, T\}$ is a point in the product space $\mathcal{Q} \times \mathcal{K} \times \mathcal{T}$, where $\mathcal{Q}$, $\mathcal{K}$ and $\mathcal{T}$ are the space of $q, k$ and $t$. Unlike the Washington algorithm, the partition is complicated and impossible to visualize.

The Daily Statistics Algorithm uses many samples (time points) of a single detector. Its main drawbacks are (1) that the day-by-day decision is too crude, and (2) the spatial correlation of good samples is not exploited. Because of (1), a moderate number of bad samples at an otherwise good detector will never be flagged. By (2), we mean that some errors that are not visible from a single detector can be readily recognized if its relationship with its spatial and temporal neighbors is considered. For example, for neighboring detectors $i$ and $j$, if the absolute difference $|q_i(t) - q_j(t)|$ is too big, either $\Delta_i = 1$ or $\Delta_j = 1$ or both. This has to do with the high lane-to-lane (and location-to-location) correlation of both $q$ and $k$. Figure 1 illustrates these points. Loops in the first and second lanes suffer from transient malfunctions, which cannot be easily detected from one-dimensional marginal distributions, but which are immediately clear from the two-dimensional joint distributions. From their relationships with lanes three and four, one can conclude that both detectors are bad.

The Washington algorithm and the DSA are ad hoc in conception, and can surely be improved upon. A systematic and principled algorithm is hard to develop mainly due to the size and complexity of the problem. An ideal detection algorithm needs to work well with thousands of detectors, all with potentially unknown types of malfunction. Even constructing a training set is not trivial since there is so much data to examine and it is not always possible to be absolutely sure if the data are correct even after careful visual inspection. [For example, suppose a detector reports $(q, k) = (0, 0)$. It could be that the detector is stuck at "off" position but good detectors will also report $(0, 0)$ when there are no vehicles in the detection period. Similarly, occupancy measurements stuck at a reasonable value will not trigger any alarm if one considers only a single detector and a single time.] New approaches should include a method of delineating acceptance/rejection regions for $k$ and $q$ for multiple sensors, combining traffic dynamics theory and manual identification of good or bad data points, with the help of interactive data analysis tools such as XGobi (http://www.research.att.com/areas/stat/xgobi/), and an intelligent way of combining evidence from various sensors to make decisions about a particular sensor/observation.

## 4. IMPUTATION

Holes in the data due to missing or bad observations must be filled with imputed values. Because of the high lane-to-lane and location-to-location correlation of $q$ and $k$, it is natural to use measurements from neighboring detectors. Although there is flexibility in the choice of a neighborhood, in practice we use the neighborhood defined by the set of loops at the same location. Let $\mathcal{N}(i)$ denote the set of neighboring detectors of $i$ and consider imputing flow, for example.

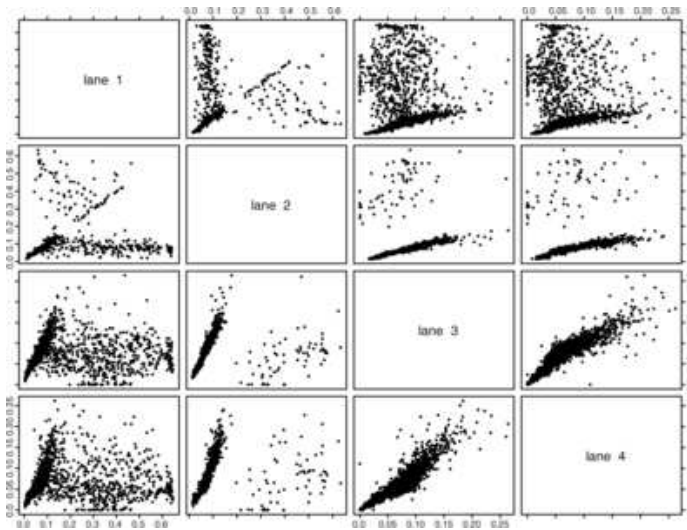

FIG. 1. *Scatter plots of occupancies at station 25 of westbound I-210.*

A natural imputation algorithm is the prediction of $q_i(t)$ based on its neighbors:

$$\hat{q}_i(t) = \hat{g}(\mathbf{q}_{\mathcal{N}(i)}(t)), \quad (2)$$

where the prediction function $\hat{g}$ is fit from historical data $\{(q_i(t), \mathbf{q}_{\mathcal{N}(i)}(t), t = 1, \ldots, T)\}$. (Note that the prediction function must be able to properly take into account possible configurations of missing and bad values among the neighbors; the latter are especially problematic, since bad readings may not be flagged as such.)

The simplest idea would be estimation by the mean

$$\hat{q}_i(d,t) = \frac{1}{\sum_{j \in \mathcal{N}(i)} 1(\hat{\Delta}_j(d,t)=0)} \cdot \sum_{j \in \mathcal{N}(i)} q_j(d,t) 1(\hat{\Delta}_j(d,t)=0)$$

or median

$$\hat{q}_i(d,t) = \operatorname{median}\{\hat{q}_j(d,t) : j \in \mathcal{N}(i), \hat{\Delta}_j(d,t) = 0\}$$

to be more robust. However, such simple interpolation is not desirable since the relationships between occupancy and flows in neighboring loops are nontrivial, that is, $q_i(t) \neq q_j(t), j \in \mathcal{N}(i)$, in general. For example, at many freeway locations, the inner lane has higher flow and lower occupancy for general free flow condition than do the outer lanes. Also, if one is close to on- or off-ramps, the relationships can be quite different.

The prediction function is rather hard to manage in its full generality because of its high dimensionality and because one does not know which values will correspond to correctly functioning detectors $[\Delta_j(t) = 0]$. From a computational point of view, the following algorithm is thus appealing:

$$(3) \quad \hat{q}_i(t) = \operatorname{average}(\hat{q}_{ij}(t) : j \in \mathcal{N}(i), \hat{\Delta}_j = 0),$$

where $\hat{q}_{ij}(t) = \hat{g}_{ij}(q_j(t))$ is the regression of $q_i(t)$ on $q_j(t)$. One computes $\hat{q}_{ij}(t)$ for all $j \in \mathcal{N}(i)$ and averages over only those values regressed on "good" neighbors. The "average" can be either mean or a robust location estimate such as the median. The latter seems preferable since all bad samples from detectors $j \in \mathcal{N}(i)$ may not be flagged.

Individual regression function $g_{ij}(q_j(t))$ can be fit in various ways. Chen et al. (2003) considered the linear regression

$$q_i(t) = \alpha_0(i,j) + \alpha_1(i,j) q_j(t) + \text{noise}$$

to produce

$$\hat{q}_{i,j}(d,t) = \alpha_0(i,j) + \alpha_1(i,j) q_j(d,t)$$

for each pair of neighbors $(i,j)$, where the parameters $\alpha_0(i,j), \alpha_1(i,j)$ are estimated by the least square using historical data. This is the approach currently being used by PeMS.

Since this approach relies upon using historical data to learn how pairs of neighboring loops behave, estimation of the regression functions must be able to cope with bad data as well. Cleaning the historical data to detect malfunctions is thus necessary, and robust estimation procedures may be preferable to least squares. We also note that an empirical Bayes perspective may be useful in jointly estimating the large set of regression functions.

## 5. ESTIMATING VELOCITY

As we have noted earlier, single-loop detectors do not directly measure velocity. This is unfortunate, because velocity is perhaps the single most useful variable for traffic control and traveller information systems. In this section we present the method currently being used to estimate velocity from single-loop data.

Let us fix a day $d$ and a time of day $t$ and consider the following situation. Suppose that at a given detector during a 30-second time interval, $N$ vehicles pass with (effective) lengths $L_1, \ldots, L_N$ and velocities $v_1, \ldots, v_N$. (The *effective vehicle length* is equal to the length of the vehicle plus the length of the loop's detector zone.) The occupancy is given by $k = \sum_{i=1}^{N} L_i / v_i$. Now, if all velocities are equal, $v = v_1 = \cdots = v_N$, it follows that

$$(4) \quad k = \frac{1}{v} \sum_{i=1}^{N} L_i = \frac{N \bar{L}}{v},$$

where $\bar{L} = \sum_{i=1}^{N} L_i / N$ is the average of the vehicle lengths. We see that if the average vehicle length is known, we can infer the common velocity. We model the lengths $L_i$ as random variables with common mean $\mu$. Note that the $L_i$ and $\bar{L}$ are not directly observed. If $\mu$ were known, while the average $\bar{L}$ is not, then a sensible estimate of the common velocity may be obtained by replacing the average by the mean in (4):

$$\hat{v} = \frac{N\mu}{k}. \tag{5}$$

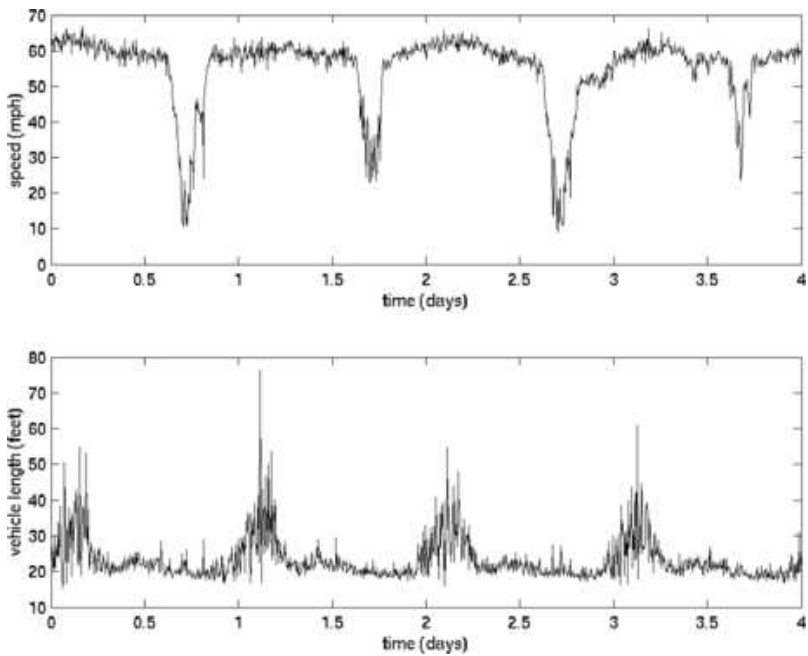

Fig. 3. *Velocity (top) and effective vehicle length (bottom) for four weekdays on I-80.*

Rewriting, we find $\hat{v} = v\mu/\bar{L}$. Since the expectation of $1/\bar{L}$ is not equal to $1/\mu$, the expectation of $\hat{v}$ is not equal to $v$. In other words, $\hat{v}$ is not an unbiased estimator of $v$, despite our assumption that all $v_i$ are equal. However, if the number of vehicles $N$ is not too small, then $\bar{L}$ should be reasonably close to its mean and the bias negligible. Henceforth, we neglect this bias issue and use formula (5) to estimate velocity. We thus focus on estimating the mean vehicle length, $\mu$.

### 5.1 Estimation of the Mean Vehicle Length

Currently, it is a widespread practice to take the mean vehicle length to be constant, independent of the time of day. The validity of this assumption has been examined by many authors (e.g., Hall and Persaud, 1989 and Pushkar et al., 1994), including ourselves (Jia et al., 2001) and it is now generally recognized that it does not generally hold. This is further illustrated by double-loop data from Interstate 80 near San Francisco, which allows direct measurement of velocity. Figure 3 shows the velocity and the average (effective) vehicle length at detector station 2 in the eastbound outer lane 5. We believe that the clear daily trend can be ascribed to the ratio of trucks to cars varying with the time of day. This is confirmed by the fact that the vehicle length in the fast lanes 1 and 2, with negligible truck presence, is almost constant. We thus assume that the mean vehicle length depends on the time of day, denote it by $\mu_t$ to reflect this dependence, and consider how $\mu_t$ can be estimated.

Suppose we have observed $N(d,t)$ and $k(d,t)$ for a number of days. Let $\alpha_{0.6}$ denote the 60th percentile of the observed occupancies. Assume that during all time intervals when $k(d,t) < \alpha_{0.6}$ all vehicles travel at a common velocity $v_{FF}$. Since we may assume that any freeway is uncongested at least 60% of the time, $v_{FF}$ may be regarded as the free flow velocity. Throughout this paper we assume that $v_{FF}$ is known or estimated from exterior sources of information.

By our assumption on constant free flow velocity, we have for all $(d,t)$ such that $k(d,t) < \alpha_{0.6}$

$$\bar{L}(d,t) = \frac{v_{FF} k(d,t)}{N(d,t)}.$$

If we assume that the average vehicle length $\bar{L}(d,t)$ does not depend on whether the occupancy is above or below the threshold, then

$$E(\bar{L}(d,t) \mid k(d,t) < \alpha_{0.6}) = E\bar{L}(d,t) = \mu_t.$$

For fixed $t$ we can obtain an unbiased estimate of $\mu_t$ as

$$\hat{\mu}_t = \frac{1}{\#\{d : k(d,t) < \alpha_{0.6}\}} \sum_{d \,:\, k(d,t) < \alpha_{0.6}} \frac{v_{FF} k(d,t)}{N(d,t)}.$$

In Figure 4 we have plotted the time of day $t$ versus $v_{FF} k(d,t)/N(d,t)$ for all times $(d,t)$ when $k(d,t) < \alpha_{0.6}$. We can now estimate the expectation $\mu_t$ of the effective vehicle length by fitting a regression line to this scatter plot, via *loess* (Cleveland, 1979). The smooth regression line seen in Figure 4 is our estimator $\hat{\mu}_t$ of $\mu_t$. Note the absence of points for times between 3 P.M. and 6 P.M. when I-80 East is always congested [$k(d,t) > \alpha_{0.6}$].

Once we have an estimator $\hat{\mu}_t$ of $\mu_t$, we define a (preliminary) estimator of $v(d,t)$ as

(6) $$\hat{v}(d,t) = \frac{N(d,t) \hat{\mu}_t}{k(d,t)}.$$

This estimator and the velocity found by the double-loop detector are plotted in Figure 5. We see that it performs very well during heavy traffic and congestion. In particular, it exhibits little bias during the time period 3 P.M. to 6 P.M. over which the smoothing shown in Figure 4 was extrapolated. Unfortunately, the variance of the estimator during times of light traffic, particularly in the early hours of each day, is unacceptably large. This is clearly visible in Figure 5 with estimated velocities on day 3 around 1 A.M. shooting up to 120 mph shortly before plummeting to 30 mph. The true velocity at that time is nearly constant at 64 mph. Recall that our preliminary estimate (6) is obtained by replacing the average (effective) vehicle length $\bar{L}(d,t)$ by (an estimate of) its expectation $\mu_t$. When only a few vehicles pass the detector during a given time interval, the average vehicle length will have a large variance. Hence, in light traffic, the average vehicle length is likely to differ substantially from the mean. For instance, if only 10 vehicles pass, then it makes a big difference if there are 6 cars and 4 trucks or 7 cars and 3 trucks. This explains the large fluctuations of our preliminary estimator $\hat{v}$ during light traffic.

### 5.2 Smoothing

Coifman (2001) suggests a simple fix for the unstable behavior of $\hat{v}$ during light traffic. He sets the estimated velocity equal to the free flow velocity $v_{FF}$ when the occupancy is low:

$$\hat{v}_{\text{coifman}}(d,t) = \begin{cases} \hat{v}(d,t), & \text{if } k(d,t) \geq \alpha_{0.6}, \\ v_{FF}, & \text{otherwise.} \end{cases}$$

The performance of this estimator, in terms of mean squared error, is certainly not bad. However, about 16 out of every 24 hours (60%), the estimated velocity is a constant and that is not realistic. We can do better, in appearance as well as in mean squared error.

It is clear that we need to smooth our preliminary estimate $\hat{v}(d,t)$, but only when the volume is small. For the purpose of real-time traffic management, it is important that our smoother be causal and easy to compute with minimal data storage. Taking all this into consideration, we used an exponential filter with varying weights. A smoothed version $\tilde{v}$ of $\hat{v}$ is defined recursively as

(7) $$\begin{aligned}\tilde{v}(d,t) &= w(d,t) \hat{v}(d,t) \\ &\quad + (1 - w(d,t)) \tilde{v}(d, t-1),\end{aligned}$$

where

(8) $$w(d,t) = \frac{N(d,t)}{N(d,t) + C},$$

and $C$ is a smoothing parameter to be specified. If the time interval is of length 5 minutes, then a reasonable value would be $C = 50$. With this value of $C$, if the volume $N(d,t)$ approaches capacity, say $N(d,t) = 100$ vehicles per 5 minutes, then there is hardly any need for smoothing and the new observation receives substantial weight 2/3. On the other hand, if the volume is very small, say $N(d,t) = 10$, then the smoothing is quite severe with the new observation receiving a weight of only 1/6.

Our filtered estimator $\tilde{v}$ is plotted in Figure 6. The correspondence with the true velocity is very good. The large variability during light traffic that plagued the preliminary estimator $\hat{v}$ has been suppressed, while its good performance during heavy traffic and congestion has been retained.

We will now explain how our filter is "inspired" by the familiar Kalman filter. Suppose that the true, unobserved velocity evolves as a simple random walk:

(9) $$v_t = v_{t-1} + \varepsilon_t, \quad \varepsilon_t \sim \mathcal{N}(0, \tau^2).$$

Suppose we observe $\hat{v}_t = N_t \hat{\mu}_t / k_t = v_t \mu_t / \bar{L}_t$, where $\hat{\mu}_t$ is our estimate of $E\bar{L}_t = \mu_t$. We will work conditionally on the observed volume $N_t$. The conditional expectation of $\hat{v}_t$ is—though not quite equal—hopefully close to $v_t$. Using a one-step Taylor approximation, we find that the conditional variance

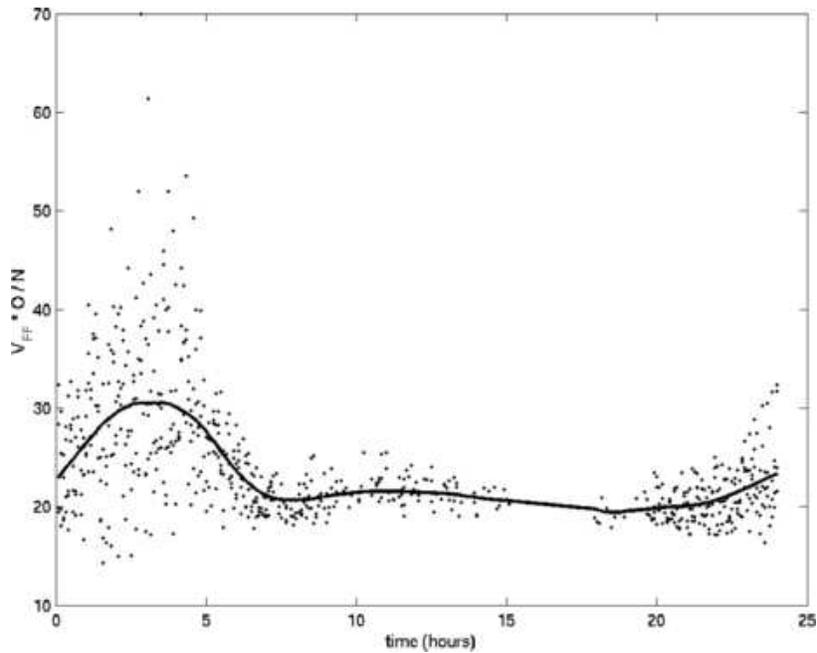

Fig. 4. *Estimation of the mean effective vehicle length $\mu_t$.*

of $\hat{v}_t$ is of the order $1/N_t$. This "inspires" a measurement equation

$$\hat{v}_t = v_t + \xi_t, \tag{10}$$
$$\xi_t \sim \mathcal{N}(0, \sigma_t^2) = \mathcal{N}(0, \sigma^2/N_t).$$

Finally, we assume that all error terms $\varepsilon_t$ and $\xi_t$ are independent. Note that the variance of the measurement error $\xi_t$ depends inversely on the observed volume $N_t$. In light traffic, when $N_t$ is small the variance is large. This is exactly the problem we noted in Figure 5.

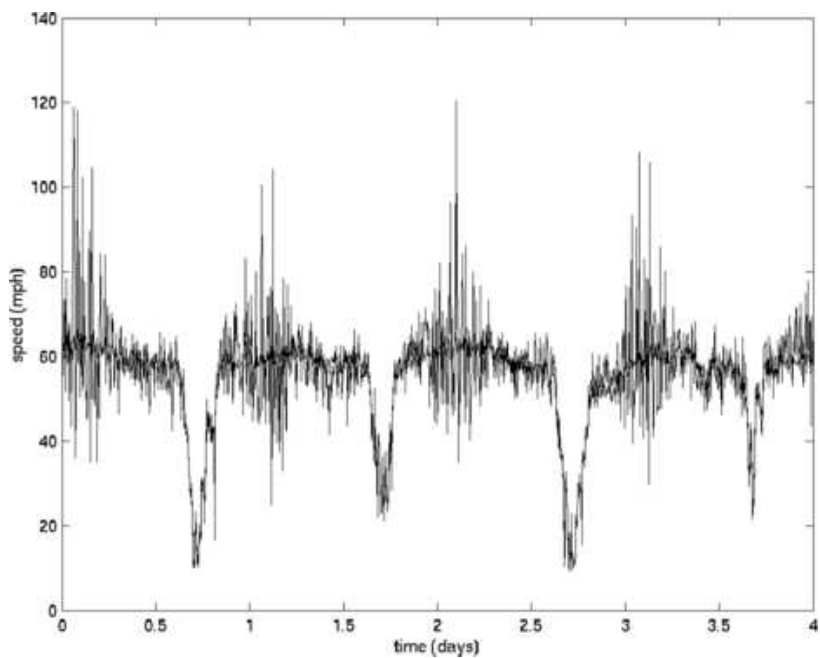

Fig. 5. *Our preliminary estimate, defined in (6), superimposed on the true velocity.*

The Kalman filter recursively computes the conditional expectation of the unobserved state variable $v_t$ given the present and past observations $\hat{v}_1, \hat{v}_2, \ldots, \hat{v}_t$:

$$\tilde{v}_t = \mathbb{E}(v_t \mid \hat{v}_1, \hat{v}_2, \ldots, \hat{v}_t).$$

In our simple model we can easily derive the Kalman recursions. They are

$$\tilde{v}_t = w_t \hat{v}_t + (1 - w_t)\tilde{v}_{t-1},$$

with

$$w_t = \frac{P_{t-1} + \tau^2}{P_{t-1} + \tau^2 + \sigma_t^2} = \frac{N_t}{N_t + \sigma^2/(P_{t-1} + \tau^2)},$$

where $P_t$ is the prediction error $\mathbb{E}(v_t - \tilde{v}_t)^2$.

We note the similarity of these Kalman recursions with our filter (7), although $C$ in (7) is constant and the analogue in the Kalman filter is not. We decided not to try to estimate $\sigma^2$ and $\tau^2$ partly because we feel that would be difficult to do reliably and partly because that would mean taking our simple model a little too seriously.

### 5.3 Known Free Flow Velocity

We assume that the free flow velocity $v_{FF}$ is known, which is typically not true. We believe that free flow velocity depends primarily on the number of lanes and on the lane number, so in practice we use values like those shown in Table 1, which are loosely based

TABLE 1
*Measured average free flow speeds (mph) for each lane (rows) of a multilane freeway depending on the total number of lanes (columns)*

| Lane number | Number of lanes | | | |
|---|---|---|---|---|
| | 2 | 3 | 4 | 5 |
| 1 | 71.3 | 71.9 | 74.8 | 76.5 |
| 2 | 65.8 | 69.7 | 71.0 | 74.0 |
| 3 | | 62.7 | 67.4 | 72.0 |
| 4 | | | 62.8 | 69.2 |
| 5 | | | | 64.5 |

on experience and empirical evidence from locations with double-loop detectors.

Clearly, it would be preferable to have an independent method to estimate site-specific free flow velocity. Petty et al.'s (1998) cross-correlation approach works well when occupancy and volume are measured in 1-second intervals. However, 20- or 30-second measurement intervals are more common and at such aggregation this method breaks down.

### 5.4 Further Assumptions on Mean Vehicle Length

We have assumed that the mean (expected) vehicle length $\mu_t$ depends on the time of day only. However, we have noticed that $\mu_t$ also depends on:

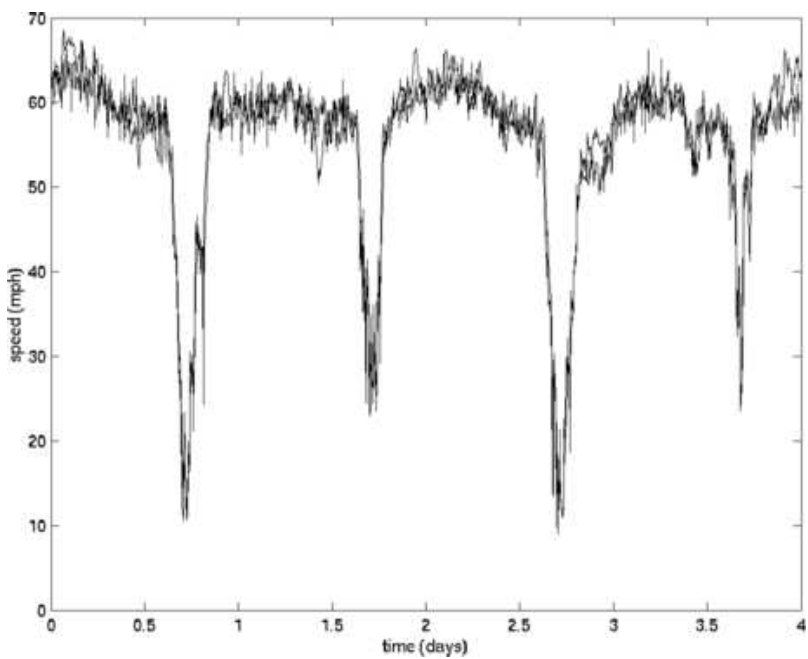

FIG. 6. *Our estimate $\tilde{V}$, defined in (7), superimposed on the true velocity.*

1. *Day of the week.* The vehicle mix on a Monday differs from a Sunday.
2. *Lane.* There is a higher fraction of trucks in the outer lanes.
3. *Location of the detector station.* Certain routes are more heavily traveled by trucks than others.
4. *Detector sensitivity.* Loop detectors are fairly crude instruments that are almost impossible to calibrate accurately. If a detector is not properly calibrated, the occupancy measurements will be biased.

To account for all this, we must form separate estimates of $\mu_t$ to cover these different situations. We store estimates of $\mu_t$ for every 5-minute interval, for every day of the week and for every lane at every detector station. In real time, the appropriate values are retrieved, multiplied by the observed volume-to-occupancy ratio and filtered.

### 5.5 Other Methods

We briefly review two other methods that also do not assume a fixed value for $\bar{L}(d,t)$, beginning with a method described in Jia et al. (2001). Suppose that we have a state variable $X(d,t)$ which is 0 during congestion and 1 during free flow. The state variable may be defined, for instance, by thresholding the occupancy $k(d,t)$. While the state is "free flow," the algorithm tracks $\bar{L}(d,t)$, assuming constant free flow velocity. As soon as the state becomes "congested," $\bar{L}(d,t)$ is kept fixed and the velocity $v(d,t)$ is tracked.

The main problem we experienced with this algorithm is that it depends crucially on $X(d,t)$. In particular, if $X(d,t) = 1$ (free flow) while congestion has already set in, the method goes badly astray. We found it difficult to develop a good rule to define $X(d,t)$. In fact, this difficulty was the main reason for us to look for a different approach.

Building on work of Dailey (1999), Wang and Nihan (2000) propose a model-based approach to estimate $\bar{L}(d,t)$ and $v(d,t)$. Their log-linear model relates $\bar{L}(d,t)$ to the expectation and variance of the occupancy $k(d,t)$, to the volume $N(d,t)$ and to two indicator functions that distinguish between high flow and low flow situations. The model has five parameters which need to be estimated from double-loop data. It is not at all clear if these parameter estimates carry over to a particular, single-loop location of interest. Wang and Nihan (2000) defer this issue to future research.

## 6. PREDICTION

We now turn our attention to travel time prediction between any two points of a freeway network for any future departure time. Regular drivers, such as commuters, choose their routes based on historical experience, but factors including daily variation in demand, environmental conditions and incidents can change traffic conditions. Since heavy congestion occurs at the time that most drivers need travel time information, free flow travel times, such as those provided by MapQuest, are of little use. The result may be inefficient use of the network. Route guidance systems based on current travel time predictions such as variable message boards could thus improve network efficiency.

We are currently developing an Internet application which will give the commuters of Caltrans District 7 (Los Angeles) the opportunity to query the prediction algorithm we describe below. The user will access our Internet site and state origin, destination and time of departure (or desired time of arrival), either using text input or interactively querying a map of the freeway system by pointing and clicking. He or she will then receive a prediction of the travel time and the best (fastest) route to take. It would also be possible to make our service available for users of cellular telephones, and in fact we plan to do so in the near future.

### 6.1 Methods of Prediction

The task is to forecast the time of a trip from loop $a$ to loop $b$ departing at some time in the future, using the information recorded up to the current time from all intervening loop detectors. One possible approach would be to model the physical process of traffic flow, using, for example a simulation program such as those mentioned in the Introduction. However, such simulations would have to be run in real time and be calibrated precisely. In general, it is not clear that the best way to predict a functional of the complex process of traffic flow is via modeling the entire process. For this reason, various purely statistical approaches, including multivariate state-space methods (Stathapoulis and Karlaftis, 2003), space–time autoregressive integrated moving average model (Kamarianakis and Prastacos, 2005), and neural networks (Dougherty and Cobbett, 1997; Van Lint and Hoogendoorn, 2002) have been proposed.

It is not obvious how to use the information from all the intervening loops, but we have found a method based on a simple compression (feature) of this data to be remarkably effective (Rice and van Zwet, 2004; Zhang and Rice, 2003). From $v$ evaluated at an array of times and loops, we can compute travel times $T_d(t)$ that should approximate the time it took to travel from loop $a$ to loop $b$ starting at time $t$ on day $d$, by "walking" through the velocity field. We can also compute a proxy for these travel times which is defined by

$$(11) \quad T_d^*(t) = \sum_{i=a}^{b-1} \frac{2u_i}{v_i(d,t) + v_{i+1}(d,t)},$$

where $u_i$ denotes the distance from loop $i$ to loop $(i+1)$. We call $T^*$ the current status travel time (a.k.a. the snap-shot or frozen field travel time). It is the travel time that would have resulted from departure from loop $a$ at time $t$ on day $d$ were there no changes in the velocity field until loop $b$ was reached. It is important to notice that the computation of $T_d^*(t)$ only requires information available at time $t$, whereas computation of $T_d(t)$ requires information at later times.

Suppose we have observed $v_l(d,t)$ for a number of days $d \in D$ in the past, that a new day $e$ has begun, and we have observed $v_l(e,t)$ at times $t \leq \tau$. We call $\tau$ the "current time." Our aim is to predict $T_e(\tau + \delta)$, the time a trip that departs from $a$ at time $\tau + \delta$ will take to reach $b$. Note that even for $\delta = 0$ this is not trivial.

Define the historical mean travel time as

$$(12) \quad \nu(t) = \frac{1}{|D|} \sum_{d \in D} T_d(t).$$

Two naive predictors of $T_e(\tau + \delta)$ are $T_e^*(\tau)$ and $\nu(\tau + \delta)$. We expect—and indeed this is confirmed by experiment—that $T_e^*(\tau)$ predicts well for small $\delta$ and $\nu(\tau + \delta)$ predicts better for large $\delta$. We aim to improve on both these predictors for all $\delta$.

6.1.1 *Linear regression.* From the extensive PeMS data, we have observed an empirical fact: that there exist linear relationships between $T^*(t)$ and $T(t+\delta)$ for all $t$ and $\delta$. This empirical finding has held up in all of numerous freeway segments in California that we have examined. It is illustrated by Figures 7 and 8, which are scatter plots of $T^*(t)$ versus $T(t+\delta)$ for a 48-mile stretch of I-10 East in Los Angeles. Note that the relation varies with the choice of $t$ and $\delta$. We thus propose the following model:

$$(13) \quad T(t+\delta) = \alpha(t,\delta) + \beta(t,\delta)T^*(t) + \varepsilon,$$

where $\varepsilon$ is a zero mean random variable modeling random fluctuations and measurement errors. Note that the parameters $\alpha$ and $\beta$ are allowed to vary with $t$ and $\delta$. Linear models with varying parameters are discussed in Hastie and Tibshirani (1993).

Fitting the model to our data is a familiar linear regression problem which we solve by weighted least squares. Define the pair $(\hat{\alpha}(t,\delta), \hat{\beta}(t,\delta))$ to minimize

$$(14) \quad \sum_{\substack{d \in D \\ s \in T}} (T_d(s) - \alpha(t,\delta) - \beta(t,\delta)T_d^*(t))^2 \cdot K(t+\delta-s),$$

where $K$ denotes the Gaussian density with mean zero and a variance which is a bandwidth parameter. The purpose of this weight function is to impose smoothness on $\alpha$ and $\beta$ as functions of $t$ and $\delta$. We assume that $\alpha$ and $\beta$ are smooth in $t$ and $\delta$ because we expect that average properties of the traffic do not change abruptly. The actual prediction of $T_e(\tau + \delta)$ becomes

$$(15) \quad \hat{T}_e(\tau + \delta) = \hat{\alpha}(\tau,\delta) + \hat{\beta}(\tau,\delta)T_e^*(\tau).$$

Writing $\alpha(t,\delta) = \alpha'(t,\delta)\nu(t+\delta)$, we see that (13) expresses a future travel time as a linear combination of the historical mean and the current status travel time, our two naive predictors. Hence our new predictor may be interpreted as the best linear combination of our naive predictors. From this point of view, we can expect our predictor to do better than both, and it does, as is demonstrated below.

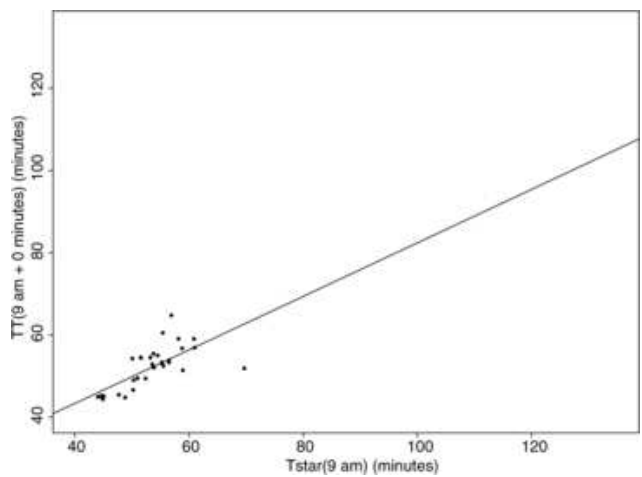

FIG. 7. $T^*$ *(9 A.M.) vs. T(9 A.M. + 0 min). Also shown is the regression line with slope $\alpha$(9 A.M., 0 min) = 0.65 and intercept $\beta$(9 A.M., 0 min) = 17.3.*

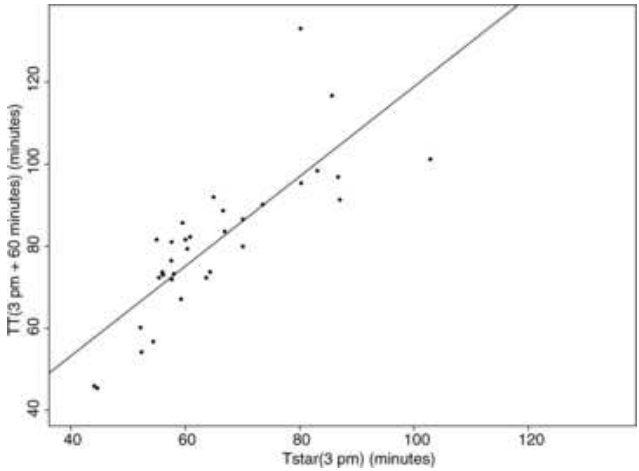

Fig. 8. $T^*(3$ P.M.$)$ vs. $T(3$ P.M. $+ 60$ min$)$. Also shown is the regression line with slope $\alpha(3$ P.M.$, 60$ min$) = 1.1$ and intercept $\beta(3$ P.M.$, 60$ min$) = 9.5$.

Another way to think about (13) is by remembering that the word "regression" arose from the phrase "regression to the mean." In our context, we would expect that if $T^*$ is much larger than average, signifying severe congestion, then congestion will probably ease during the course of the trip. On the other hand, if $T^*$ is much smaller than average, congestion is unusually light and the situation will probably worsen during the journey.

In addition to comparing our predictor to the historical mean and the current status travel time, we subject it to a more competitive test. We consider two other predictors that may be expected to do well, one resulting from principal component analysis and one from the nearest-neighbors principle. Next, we describe these two methods.

6.1.2 *Principal components.* Our predictor $\hat{T}$ only uses information at one time point: the "current time" $\tau$. However, we do have information prior to that time. The following method attempts to exploit this by using the entire trajectories of $T_e$ and $T_e^*$ which are known up to time $\tau$.

Formally, let us assume that the travel times on different days are independently and identically distributed and that for a given day $d$, $\{T_d(t):t \in T\}$ and $\{T_d^*(t):t \in T\}$ are jointly multivariate normal. We estimate the large covariance matrix of this multivariate normal distribution by retaining only a few of the largest eigenvalues in the singular value decomposition of the empirical covariance of $\{(T_d(t), T_d^*(t)):d \in D, t \in T\}$. Define $t'$ to be the largest $t$ such that $t + T_e(t) \leq \tau$. That is, $t'$ is the (random) start time of the latest trip that we would have seen completed if we observed day $d$ until time $\tau$. With the estimated covariance we can now compute the conditional expectation of $T_e(\tau + \delta)$ given $\{T_e(t):t \leq t'\}$ and $\{T_e^*(t):t \leq \tau\}$. This is a standard computation which is described, for instance, in Mardia et al. (1979). The resulting predictor is $\hat{T}_e^{\mathrm{PC}}(\tau + \delta)$.

6.1.3 *Nearest neighbors.* As an alternative, we now consider another attempt to use information prior to the current time $\tau$, based on nearest neighbors. This nonparametric method makes fewer assumptions (such as joint normality) on the relation between $T^*$ and $T$ than does the principal components method, but is tied to a particular metric.

The nearest-neighbor method uses that day in the past which is most similar to the present day in some appropriate sense. The remainder of that past day beyond time $\tau$ is then taken as a predictor of the remainder of the present day.

The method requires a suitable distance $m$ between days. We have investigated two possible distances:

$$(16) \quad m_1(e,d) = \sum_{i=a,\ldots,b, t \leq \tau} |v_i(e,t) - v_i(d,t)|$$

and

$$(17) \quad m_2(e,d) = \left(\sum_{t \leq \tau}(T_e^*(t) - T_d^*(t))^2\right)^{1/2}.$$

Now, if day $d'$ minimizes the distance to $e$ among all $d \in D$, our prediction is

$$(18) \quad \hat{T}_e^{NN}(\tau + \delta) = T_{d'}(\tau + \delta).$$

Sensible modifications of the method are windowed nearest neighbors and $k$-nearest neighbors. Windowed-NN recognizes that not all information prior to $\tau$ is equally relevant. Choosing a window size $w$, it takes the above summation to range over all $t$ between $\tau - w$ and $\tau$. The $k$-nearest neighbor modification finds the $k$ closest days in $D$ and bases a prediction on a (possibly weighted) combination of these. However, neither of these variants appears to significantly improve on the vanilla $\hat{T}^{NN}$.

**6.2 Results**

To compare these methods we used flow and occupancy data from 116 single-loop detectors along 48 miles of I-10 East in Los Angeles (between postmiles 1.28 and 48.525). Measurements were done at 5-minute aggregation at times $t$ ranging from 5 A.M. to 9 P.M. for 34 weekdays between June 16 and September 8, 2000. We used the methods we have previously described to convert flow and occupancy to velocity.

The quality of our I-10 data is quite good and we have used simple interpolation to impute wrong or missing values. The resulting velocity field $v_i(d,t)$ is shown in Figure 9 where day $d$ is June 16. The horizontal streaks typically indicate detector malfunction.

From the velocities we computed travel times for trips starting between 5 A.M. and 8 P.M. Figure 10 shows these $T_d(t)$ where time of day $t$ is on the horizontal axis. Note the distinctive morning and afternoon congestions and the huge variability of travel times, especially during those periods. During afternoon rush hour we find travel times of 45 minutes to up to two hours. Included in the data are holidays July 3 and 4 which may readily be recognized by their very short travel times.

We have estimated the root mean squared (RMS) error of our various prediction methods for a number of "current times" $\tau$ ($\tau = 6$ A.M., 7 A.M.,..., 7 P.M.) and lags $\delta$ ($\delta = 0$ and 60 minutes). The RMS errors were estimated by leaving out one day at a time, performing the prediction for that day on the basis of the remaining other days, and averaging the squared prediction errors.

The prediction methods all have smoothing parameters that must be specified. For the regression method we chose the standard deviation of the Gaussian kernel $K$ to be 10 minutes. For the principal components method we chose the number of eigenvalues retained to be four. For the nearest-neighbors method we have chosen distance function (17), a window $w$ of 20 minutes and the number $k$ of nearest neighbors to be two. The results were fairly insensitive to these precise choices.

Figures 11 and 13 show the estimated RMS prediction errors of the historical mean $\nu(\tau + \delta)$, the current status predictor $T_e^*(\tau)$ and our regression predictor (15) for lag $\delta$ equal to 0 and 60 minutes, respectively. Note how $T_e^*(\tau)$ performs well for small $\delta$ ($\delta = 0$) and how the historical mean does not become worse as $\delta$ increases. Most importantly, however, notice how the regression predictor dominates both.

Figures 12 and 14 again show the RMS prediction error of the regression estimator. This time, it is compared to the principal components predictor and the nearest-neighbors predictor (18). Again, the regression predictor comes out on top, although the nearest-neighbors predictor shows comparable performance.

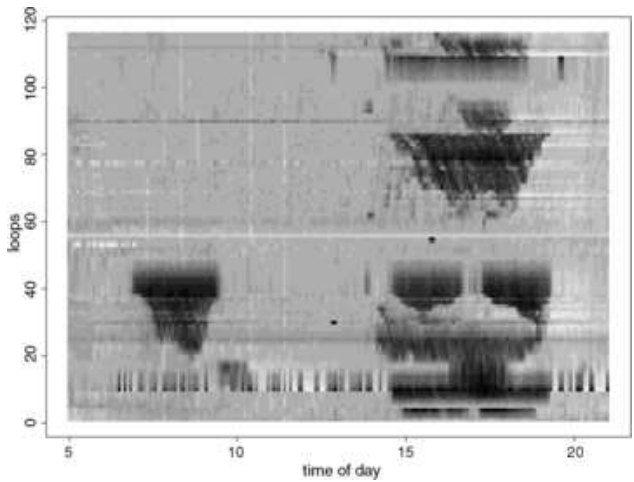

FIG. 9. *Velocity field $V(d,l,t)$ where day $d =$ June 16, 2000. Darker shades indicate lower speeds.*

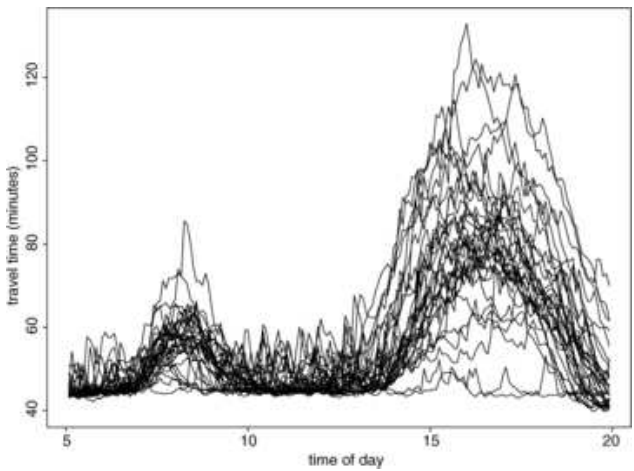

FIG. 10. *Travel times $T_d(\cdot)$ for 34 days on a 48-mile stretch of I-10 East.*

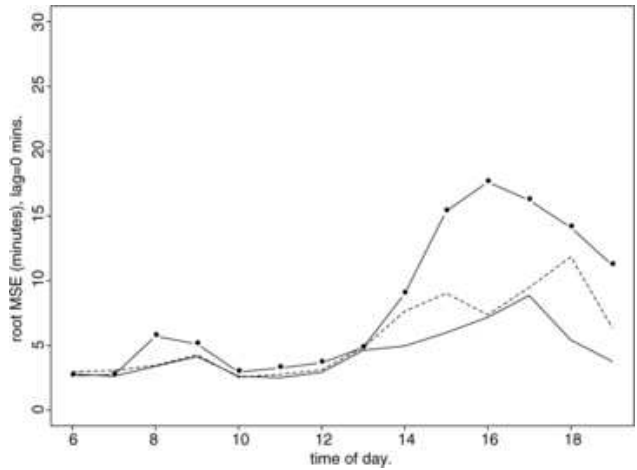

FIG. 11. *Estimated RMSE, lag $= 0$ minutes. Historical mean $(-\cdot-)$, current status (- - -) and linear regression (—).*

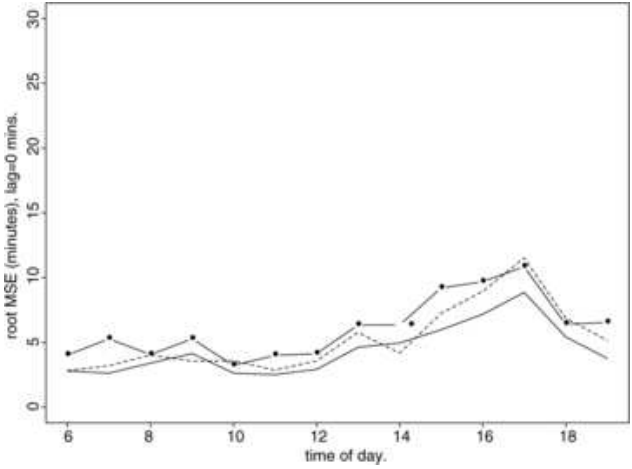

FIG. 12. *Estimated RMSE, lag = 0 minutes. Principal components (– · –), nearest neighbors (- - -) and linear regression (—).*

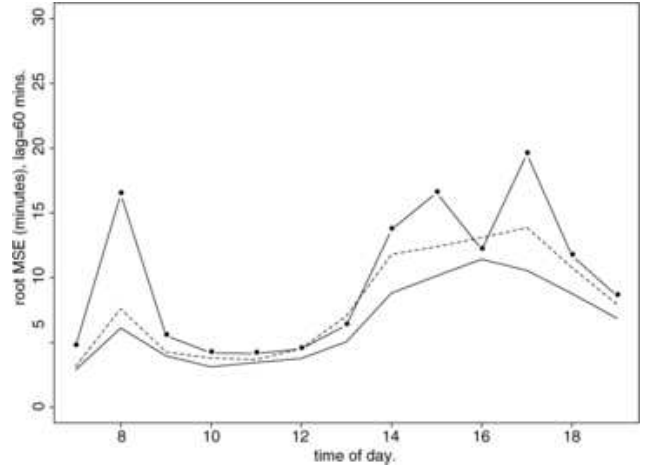

FIG. 14. *Estimated RMSE, lag = 60 minutes. Principal components (– · –), nearest neighbors (- - -) and linear regression (—).*

The RMS error of the regression predictor stays below 10 minutes even when predicting an hour ahead. We feel that this is impressive for a trip of 48 miles through the heart of Los Angeles during rush hour.

Comparison of the regression predictor to the principal components and nearest-neighbors predictors is surprising: the results indicate that given $T^*(\tau)$, there is not much information left in the earlier $T^*(t)$ ($t < \tau$) that is useful for predicting $T(\tau + \delta)$, at least by the methods we have considered. In fact, we have come to believe that for the purpose of predicting travel times, all the information in the $v_l(d,t)$ up to time $\tau$ is well summarized by one single number: $T^*(\tau)$.

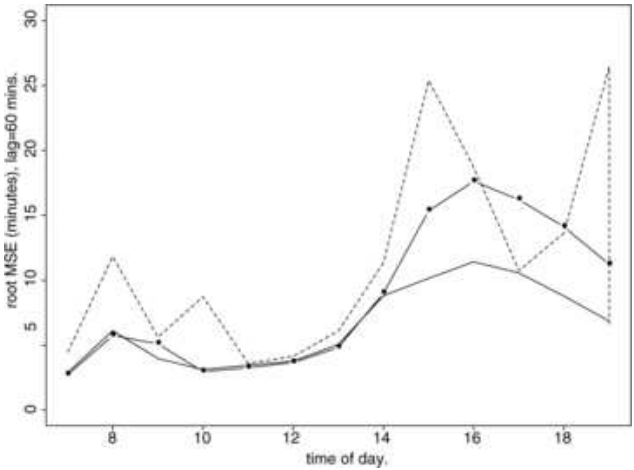

FIG. 13. *Estimated RMSE, lag = 60 minutes. Historical mean (– · –), current status (- - -) and linear regression (—).*

Recently, Nikovski et al. (2005) compared the performance of several statistical methods on data from a 15-km stretch of freeway in Japan. Their conclusions mirrored ours: a regression approach outperformed neural networks, regression trees and nearest-neighbor methods. They also reached the conclusion that the predictive information is contained in the current travel time.

### 6.3 Further Remarks

It is of practical importance to note that our prediction can be performed in real time. Computation of the parameters $\hat{\alpha}$ and $\hat{\beta}$ is time consuming but it can be done off-line in reasonable time. The actual prediction is then trivial to compute.

We conclude this section by briefly pointing out two extensions of our prediction method:

1. For trips from $a$ to $c$ via $b$ we have

$$(19) \quad T_d(a,c,t) = T_d(a,b,t) + T_d(b,c,t + T_d(a,b,t)).$$

We have found that it is sometimes more practical or advantageous to predict the terms on the right-hand side than to predict $T_d(a,c,t)$ directly. For instance, when predicting travel times across networks (graphs), we need only predict travel times for the edges and then use (19) to piece these together to obtain predictions for arbitrary routes.

2. In the discussion above we regressed the travel time $T_d(t+\delta)$ on the current status $T_d^*(t)$, where $T_d(t+\delta)$ is the travel time departing at time $t+\delta$. Now, define $S_d(t)$ to be the travel time *arriving* at time $t$ on day $d$. Regressing $S_d(t+\delta)$ on $T_d^*(t)$ allows us to make predictions on the travel time subject to *arrival* at time $t+\delta$. The user can thus ask what time he or she should depart in order to reach an intended destination at a desired time.

## 7. CONCLUSION

Modern communication and computational facilities make possible, in principle, systematic use of the vast quantities of historical and real-time data collected by traffic management centers. Such efforts invariably require substantial use of statistical methodology, often of a nonstandard variety, sensitive to computational efficiency.

This paper has concentrated on data collected by inductance loops in freeways, but similar data is often available on arterial streets as well, which have more complex flows and geometry. There is also information from other types of sensors. For example, declining costs make video monitoring an attractive technology, bringing with it challenging problems in computer vision and statistics. As another example, data derived from transponders installed in individual vehicles for automatic toll payments is a potentially rich source of information about traffic flow, since the tags can in principle be sensed at locations other than toll booths. Effective extraction of information will require active collaborations of statisticians, traffic engineers, and specialists in various other disciplines.

## ACKNOWLEDGMENTS

This study is part of the PeMS project, which is supported by grants from Caltrans to the California PATH Program. We are very grateful to Caltrans Traffic Operations engineers for their support. Our research has also been supported in part by grants from the National Science Foundation.

The contents of this paper reflect the views of the authors, who are responsible for the facts and the accuracy of the data presented herein. The contents do not necessarily reflect the official views of or policy of the California Department of Transportation. This paper does not constitute a standard, specification or regulation.